**Neuron-Inspired Flexible Memristive Device on Silicon (100)**

*Mohamed T. Ghoneim, Muhammad M. Hussain*

Integrated Nanotechnology Lab, Electrical Engineering, Computer Electrical Mathematical Science and Engineering Division, King Abdullah University of Science and Technology, Thuwal 23955-6900, Saudi Arabia

## Abstract

Comprehensive understanding of the world's most energy efficient powerful computer – human brain is an illusive scientific issue. Still, already gained knowledge indicates memristors can be used as a building block to model the brain. At the same time, brain cortex is folded allowing trillions of neurons to be integrated in a compact area. Therefore, we report flexible aluminum oxide based memristive device fabricated and then derived from widely used bulk mono-crystalline silicon (100). We use complementary-metal-oxide-semiconductor based processes to layout the foundation for ultra-large-scale-integration (ULSI) of such memory devices to advance the task of comprehending a physical model of human brain.

**Keywords:** Brain, neuron, memristor, resistive, flexible, silicon.

Memristors are the fourth missing fundamental circuit component. Although theoretically envisioned by L. Chua in 1971,[1] the concept has been revived in 2008 when R. Stanley Williams *et al.* of HP labs have experimentally demonstrated Chua's missing memristor via a simple Platinum(Pt)/Titanium Oxide ($TiO_2$)/Pt structure.[2] Since then memristor devices structures and memristive material systems have been investigated due to their scalability down to regimes beyond status-quo enabled by their relatively simple structure.[3] Memristors show a great potential of replacing not only flash type non-volatile memory element but also volatile but high performance dynamic random access memory (DRAM) and static random access memory (SRAM).[4-8] This great potential is due to the desired attributed of resistive switching in general such as low power, fast switching and long retention time. To develop a reliable memristor device, different transitional metal oxides and binary compounds have been studied, such as Nickel Oxide ($NiO_x$), $TiO_2$ and Aluminum Oxide ($Al_2O_3$).[9] From manufacturability perspective, $NiO_x$ and $TiO_2$ offer limited opportunity as being very hard materials, their patterning required heavy ions based non-selective physical sputtering based reactive ion etching leaving irrecoverable damage and potential plasma induced damage in the underlying material(s).[10,11] On the other hand, $Al_2O_3$ can be etched using reactive ion etching with $CHF_3$ which is selective to most of the common materials used in the semiconductor industry.[12] Furthermore, $Al_2O_3$ is a high-κ dielectric material ($\varepsilon_r \sim 9$) commonly used as a thermally stable gate dielectric and can be formed by versatile techniques including thermal oxidation of Al, sputtering, or atomic layer deposition (ALD). $Al_2O_3$ memristors have been previously reported to exhibit unipolar[13] as well as bipolar resistive switching.[14-15] The reported works on $Al_2O_3$ is widely varied in terms of endurances ranging from few hundred cycles[16] to thousands of cycles[15] and a high resistance state (HRS) to low resistance state (LRS) with a ratio of few[17] to thousands.[15] Different variations make these devices more versatile and tuning the stoichiometry, thickness, and phase of the material are keys in determining its resistive behavior.

In the recent past, understanding the world's most energy efficient powerful computer human brain has gained momentum and often memristor has been linked as a potential building block analog to brain neuron.[18] At the same time, brain cortex is folded which allows integration of trillions of neurons in an ultra-compact arrangement. Therefore, it is important to explore possibility for ultra-large-scale-integration (ULSI) of memristor devices on flexible platform. Although previous attempts to demonstrate flexible memory (not from the perspective of physical model of brain though), they have either used naturally flexible polymeric substrates limiting ULSI and thermal budget (impacting device performance) or hybrid approach to integrate transfer based silicon (for high performance) with polymeric substrates – not addressing the challenges with limited integration density and thermal budget.[19,20] Although it is possible to use industry regarded back grinding, such process result in limited bendability (correlated to thicker substrate of 0.1 mm), pushing the boundary lower than standard 0.1 mm using back grinding results in defect formation and device damage. Other approaches include spalling (potentially including expensive high energy implantation) and anodic etching followed by epitaxial growth – both of these processes results in limited flexibility, opaqueness and the most importantly higher cost.

In this work we report a complementary-metal-oxide-semiconductor (CMOS) compatible low-cost (45 cents/cm$^2$ additional cost) process using trench-protect-release-recycle[21-24] to demonstrate $Al_2O_3$ based memristors on flexible silicon (100). Our approach is complementary to our previous reports and is aimed at providing a facile method of achieving flexible memristive devices on the most widely used silicon (100) substrates. Our approach enables utilizing the inherent capabilities and well established infrastructure of the semiconductor industry while adding the flexibility feature. The flexible $Al_2O_3$ memristive devices on silicon exhibit hybrid complementary resistive switching (CRS) and memristive switching. We assess the performance of such devices in terms of cycle-to-cycle and device-to-device variation, and endurance against switching pulses and readout time relative to their identical bulk counterparts.

We followed a release first approach for building the memristors by releasing a thin (25 μm thick) silicon fabric then we fabricated the devices. A 4'' bulk Si wafer with 300 nm thermal $SiO_2$ was patterned to create a network of 10 μm diameter circular holes separated by 10 μm distance (edge-to-edge) in the oxide using AZ ECI 3027 photoresist and oxide reactive ion etching (RIE) using $CF_4$ and $CHF_3$ gases. Then deep RIE (DRIE) of 25 μm is silicon was performed using the Bosch process of successive etch/deposit cycles using $SF_6$ and $C_4F_8$, respectively. Then, 50 nm $Al_2O_3$ was deposited at 300 °C using followed by anisotropic reactive ion etching using $CF_4$ and $CHF_3$, leaving sidewalls (spacers) of the deep trenches protecting the bulk silicon (sideways). Finally, xenon difluoride ($XeF_2$) was used to isotropically etch the silicon from the bottom end of the deep cylindrical trenches. Due to isotropic nature of the etch process, silicon is used forming scallop shaped voids and when adjacent trenches are merged, top portion of the silicon is released as a thin sheet of silicon fabric.

The thin silicon fabric is used as our flexible platform on which the memristive devices are built on. The released silicon fabric stays anchored from all four edges to the rest of the bulk silicon (100) wafer until all the devices are fabricated. Devices are fabricated by physical vapor deposition (sputtering) of 200 nm Al based common bottom electrode, followed by a 20 nm Tantalum Nitride (TaN)/10nm $Al_2O_3$/20nm TaN atomic layer deposited stack. The TaN layer enables maintaining the quality of the ALD $Al_2O_3$. The top aluminum electrode is deposited by sputtering another 200 nm of aluminum then patterned using metal RIE ($Cl_2$:$BCl_3$:Ar—4:1:1). The patterned aluminum top electrode is then used as hard mask for nitride and oxide etching, using ($SF_6$ and $CHF_3$:Ar—4:1, respectively) respectively, of the TaN/$Al_2O_3$/TaN stack to gain access to the common bottom electrode. Figure 1 summarizes the fabrication process.

The memristive devices were characterized using Keithley 4200 semiconductor characterization system. Figure 2 shows the IV plots for unreleased (the devices which have not gone through any trench-protect-release-recycle process) and released devices over the first 5 cycles. The observable variation in figure 2(b) between the 1st and the 5th cycle of the released memristors is attributed to the forming process of the device. In accordance with the previous reports such device forming can be done by several cycles until conformity is reached or a single cycle at a much higher voltage.[25] However, since the voltage range is already too high we pursued the cycles forming approach to avoid physically damaging the devices. The sweeping cycles were done following a DC sweep 0→11→0→-11→0 V.

Figure 3 shows the direction of sweeping. In the first portion of the sweep (0→11 V), the device is in the high resistance state (HRS) until it experiences a form of soft breakdown and switches to the low resistance state (LRS) at 11 V. Then the device stays at the LRS until it reaches -11 V then switches back to HRS. The non-zero crossing is attributed to the capacitive effects of the devices due to the large size (100 x 100 μm$^2$).[26] This aspect can be useful in applications where complementary resistive switching is required as OFF state leakage is significantly low.[27]

Figure 4 shows the device to device variation of 10 devices depicting similar behavior for unreleased as well as released devices. Finally, Figure 5 shows the variation in HRS and LRS of the memristive devices as more switching pulses are applied as well as when the read voltage is applied over extended periods of time. Switching pulses corresponding to HRS and LRS were -11 V and 11 V, respectively. The read voltage was 7 V. The high set, reset and read voltages might cause severe degradation in the devices leading to fast deterioration as evident in successive endurance tests. The endurance plots (Figure 5) show the severe degradation in the ratio of HRS/LRS which is already low (3-4 times) to start with and failure to operate as the ratio approaches 1 after ~110 pulses and 80 pulses for unreleased and released devices, respectively, exhibiting 27% degradation. Although the low RHS/LRS ratio is useful for analog circuits and applications the low number of pulses possesses a challenge. It has been previously reported that $Al_2O_3$ memristive behavior can withstand only few hundreds of pulses,[16] as well as other reporting of 10$^3$ cycles.[15] Hence, endurance can be improved by further optimization of the materials as we are mainly interested in preserving the memristive aspects of flexible devices using trench-protect-release-reuse process. As for time endurance both released and unreleased devices experience a 60% decrease in the HRS value after 30 minutes while the LRS remains almost constant. This confirms the previous reporting that the HRS conduction is dominated by stress induced defects generation (similar to SILC behavior) transport while the low resistance state is insensitive to the continuously applied stress.[28]

We have demonstrated a facile CMOS compatible low-cost fabrication process for building flexible memristive devices on ultra-thin silicon fabric released from bulk mono-crystalline silicon (100). The fabricated devices on flexible silicon fabric show unperturbed performance compared to their identical bulk counterparts built on the same wafer. The only degradation observed was 27% degradation in endurance versus number of pulse switching. This work presents a foundation to achieve ultra-compact physical modeling of human brain using memristor as building block analog to brain neurons. Usage of silicon (compared to polymeric substrate) will allow ULSI of ultra-dense memory cells.

**E-mail of the corresponding author:** muhammadmustafa.hussain@kaust.edu.sa

**Acknowledgement**
We acknowledge the financial support under KAUST Office of Competitive Research Grants CRG-1 Award (CRG-1-2012-HUS-008) for this work.

Labels: Si wafer | SiO$_2$ | TaN | Al$_2$O$_3$ | Al

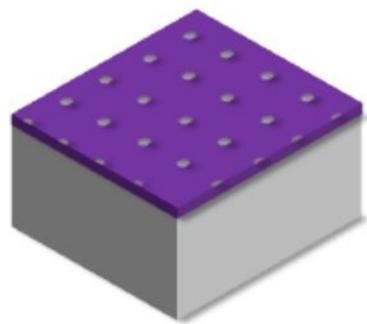

1. Oxidation and Patterning

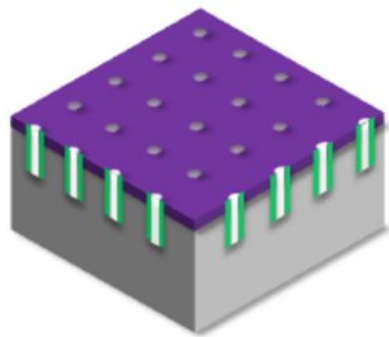

2. DRIE and Oxidation

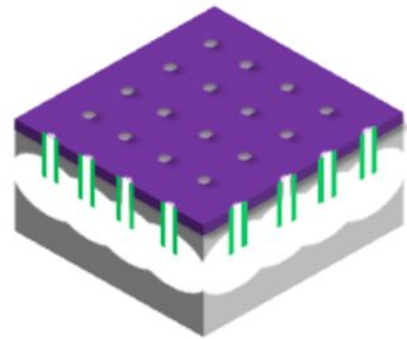

3. Release with XeF$_2$

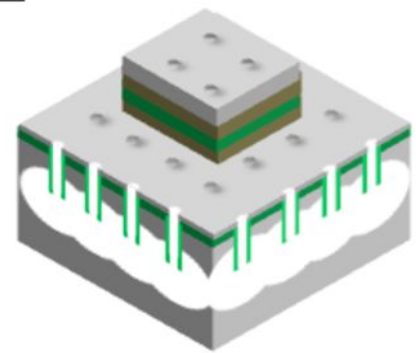

4. MIMCAP Stack Deposition and Patterning

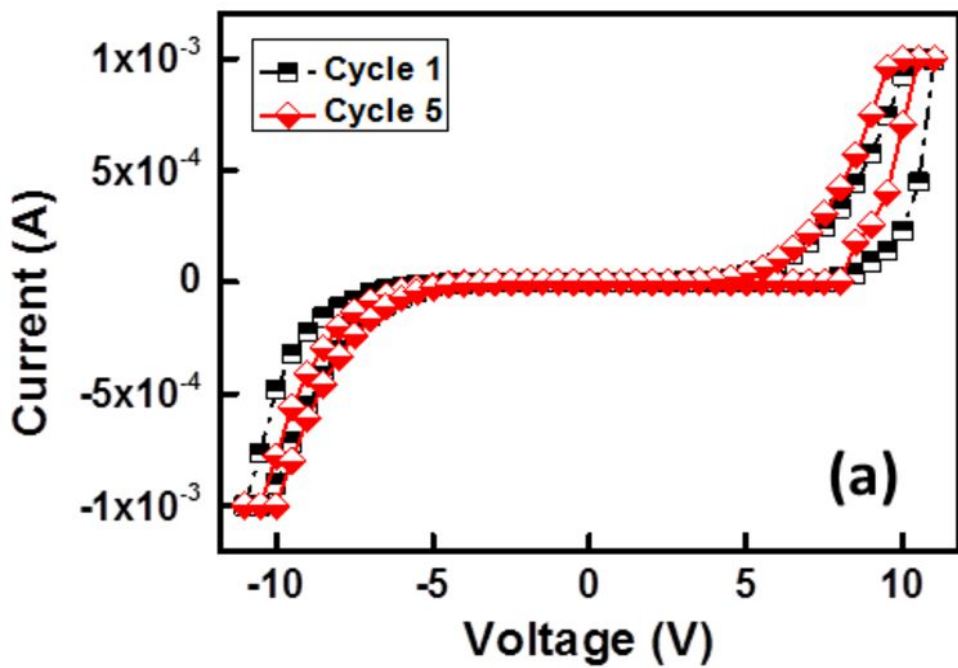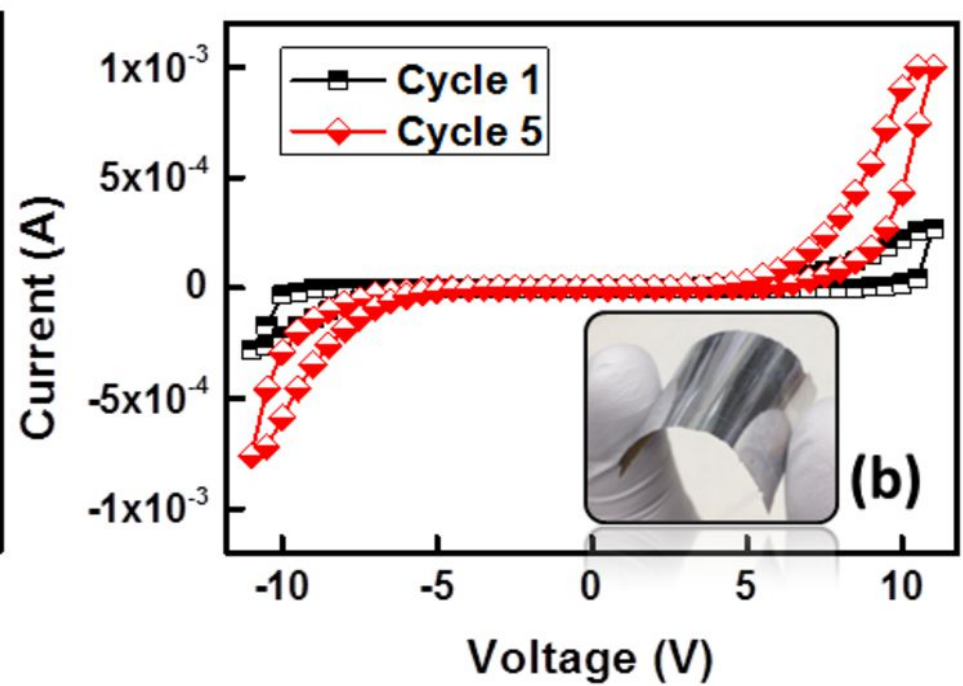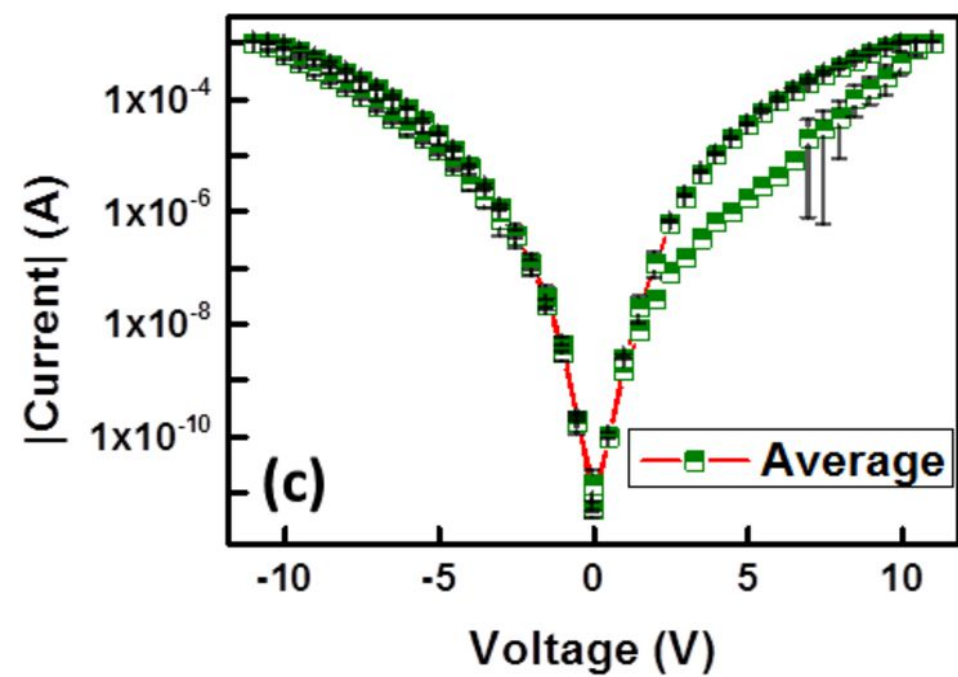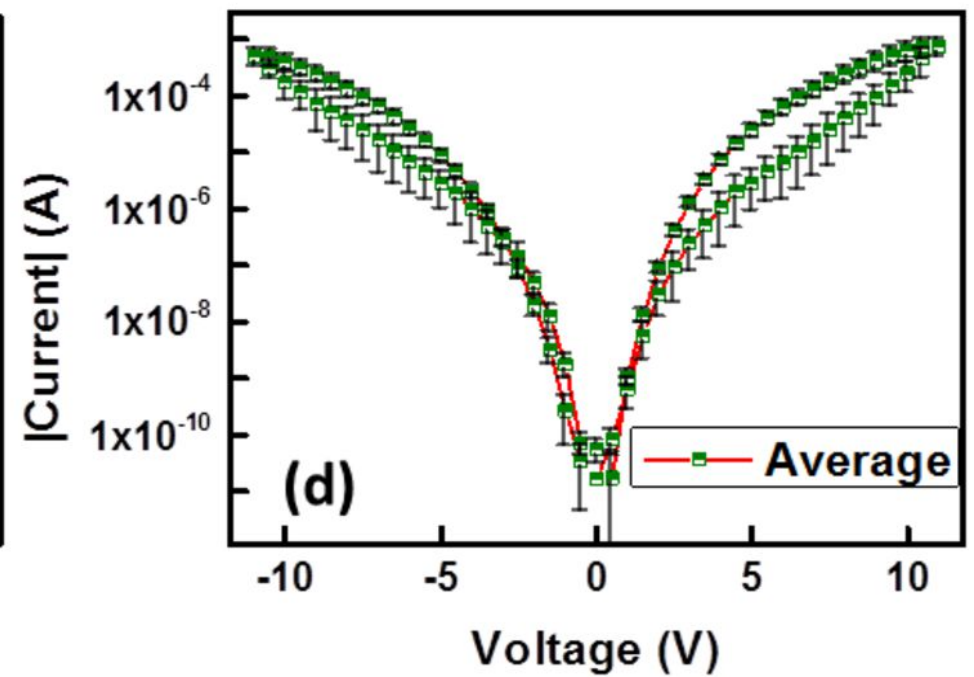

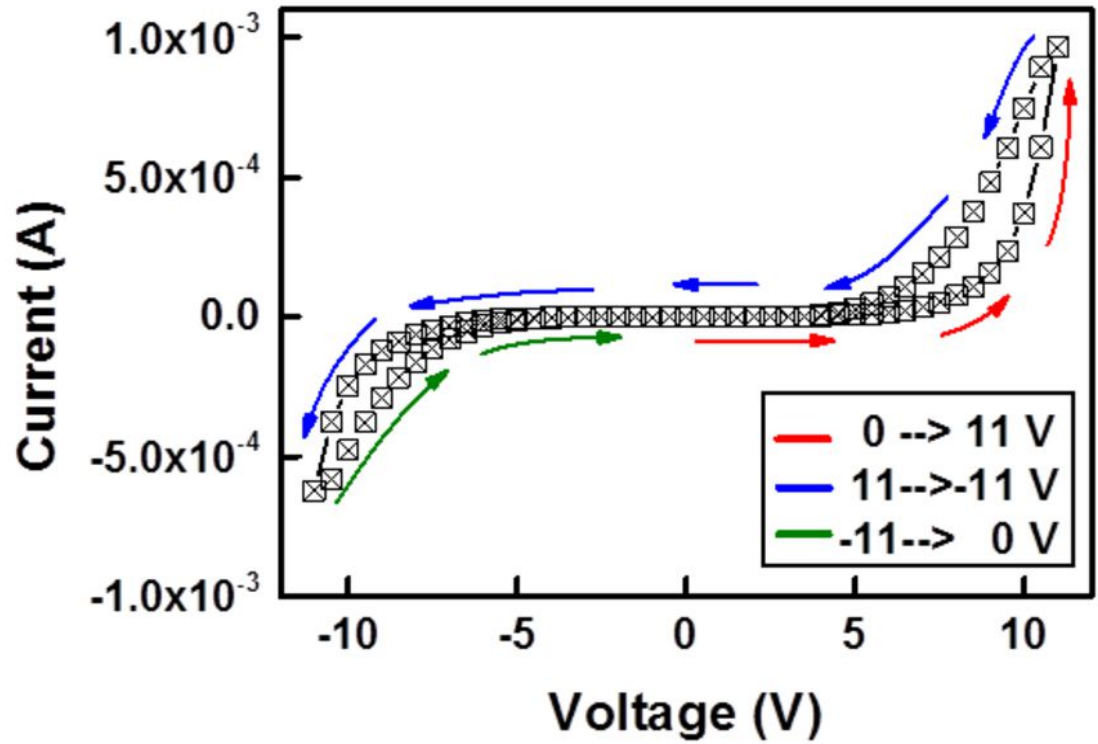

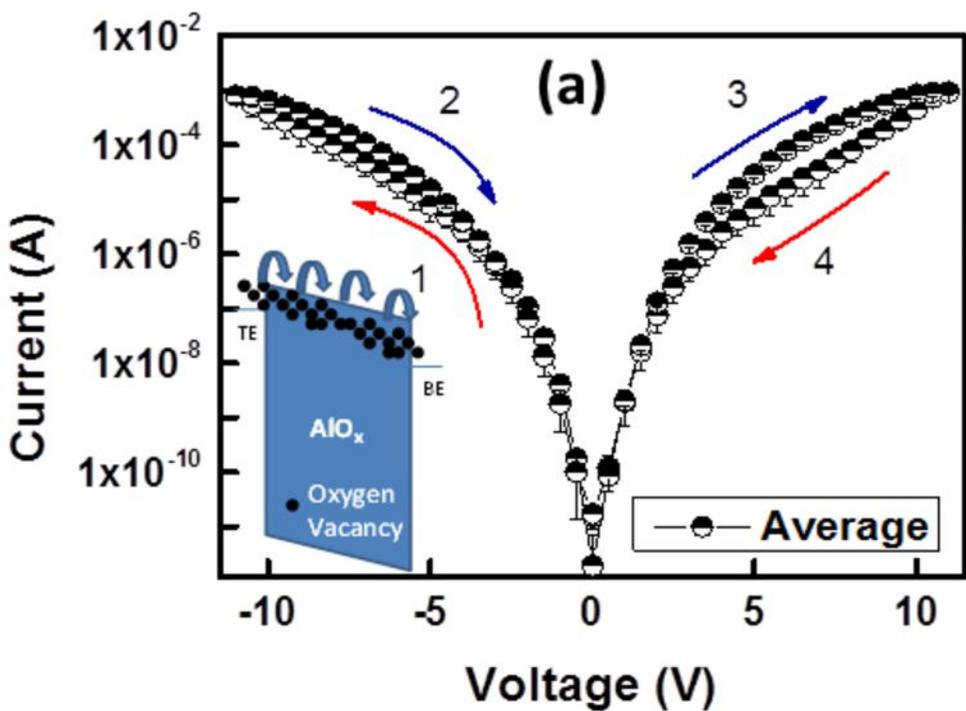 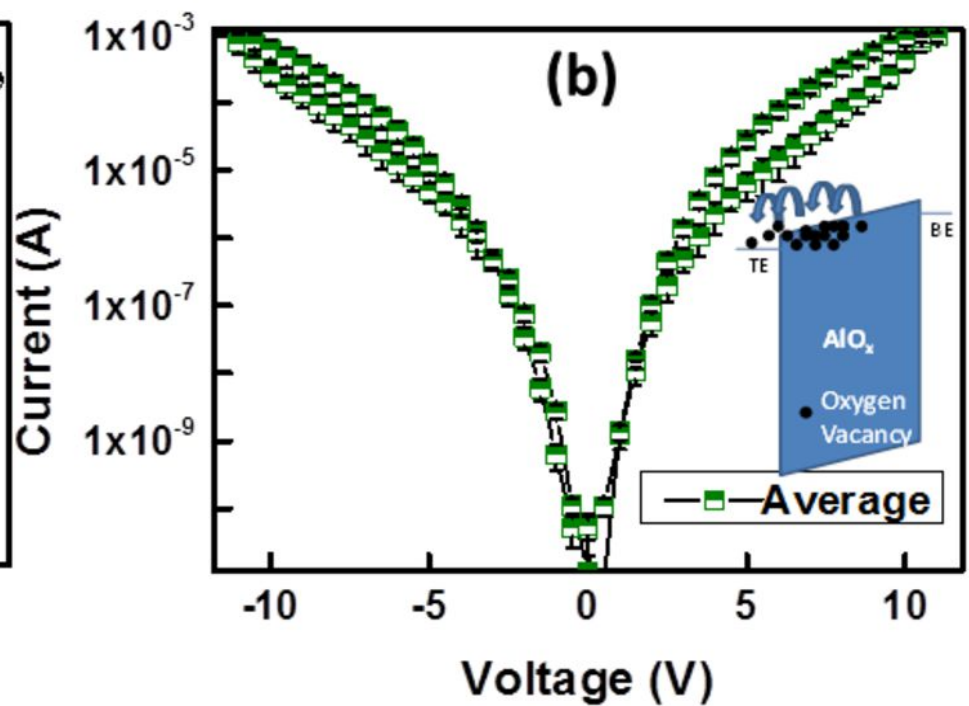

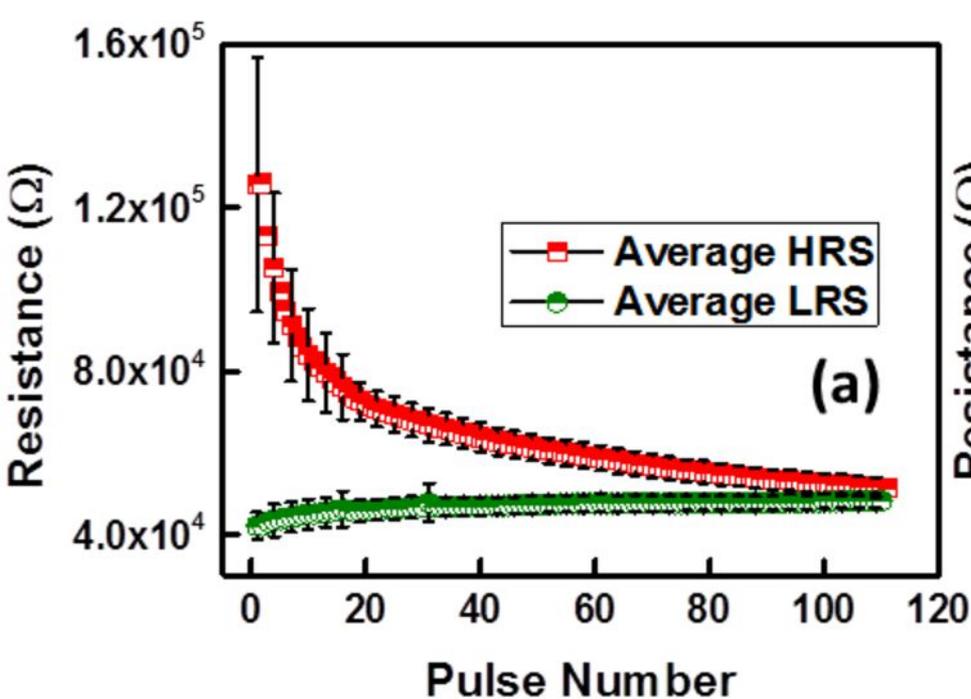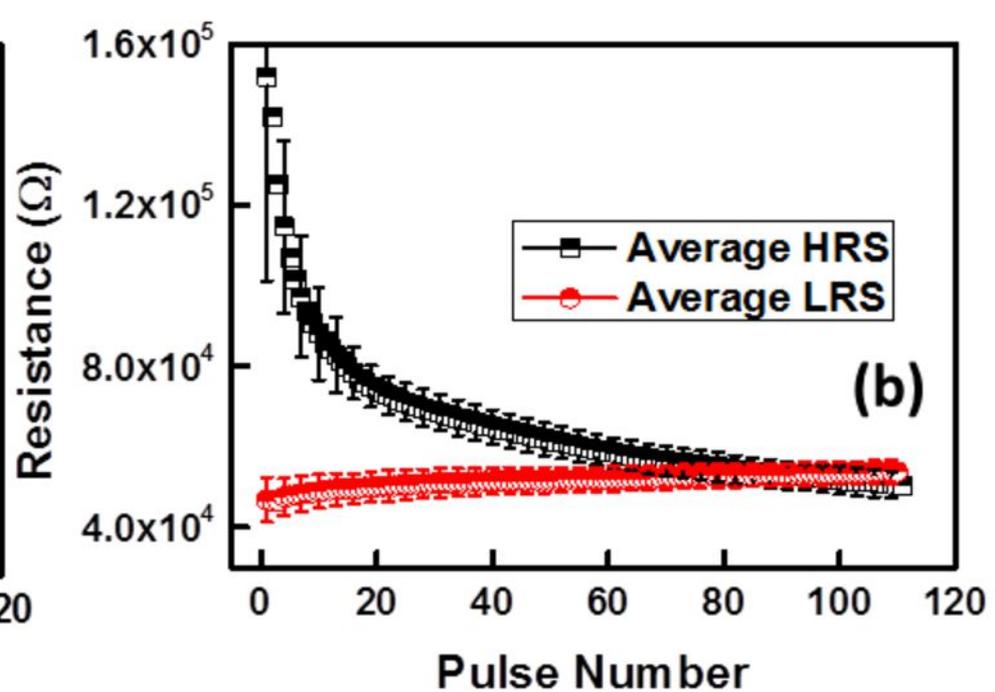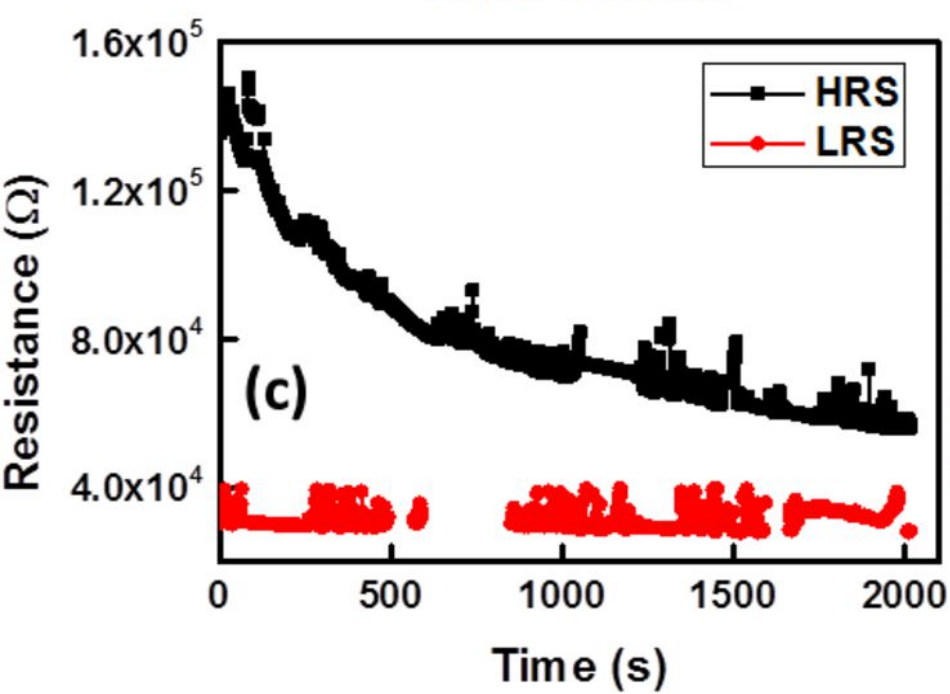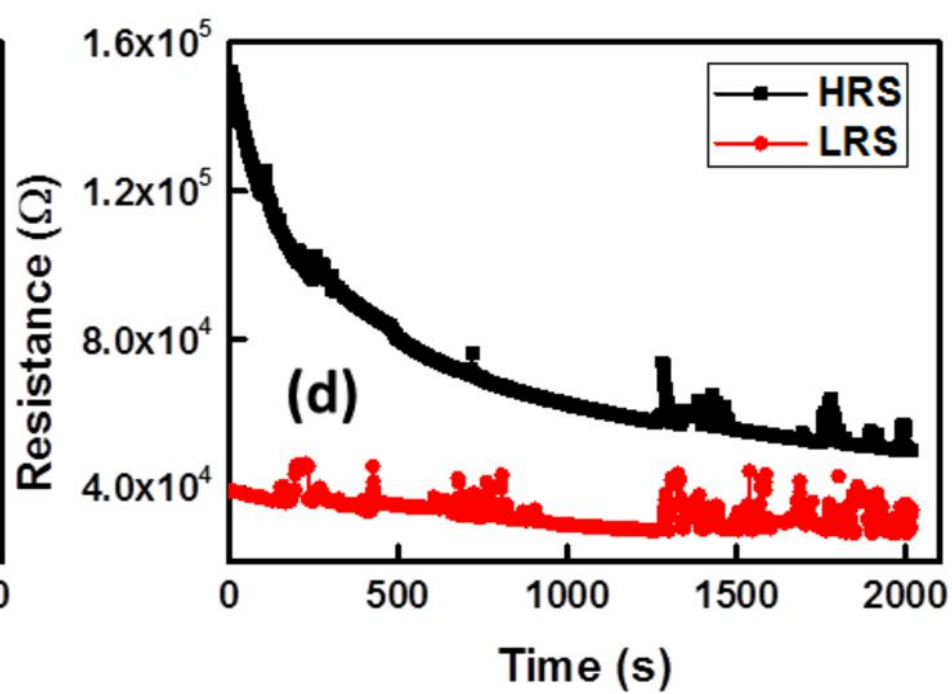